\documentstyle[11pt,aaspp4]{article}

\slugcomment{To appear in {\it Geophys. Res. Lett.}}

\begin{document}
\title{Temporal and Spectral Properties of Gamma-Ray Flashes} 
\author{Hua Feng\altaffilmark{1},
T. P. Li\altaffilmark{2,3},
Mei Wu\altaffilmark{2},
Min Zha\altaffilmark{2},
and Q. Q. Zhu\altaffilmark{2}}
\altaffiltext{1}
{Department of Engineering Physics, Tsinghua University,
Beijing, China.}

\altaffiltext{2}
{Institute of High Energy Physics, Chinese Academy of Sciences, Beijing, China.}

\altaffiltext{3}
{Physics Department and Astrophysics Center, Tsinghua University, Beijing, China.}

\begin{abstract}
The temporal and spectral properties of terrestrial gamma-ray
flashes (TGFs) are studied. The delay of low energy photons
relative to high energy ones in the $\gamma$-ray variations of
the TGFs with high signal to noise ratio has been revealed
by an approach of correlation analysis in the time domain on
different time scales. The temporal structures of the TGFs in
high energy band are usually narrower than that in low energy band. The
spectral hardness has a general trend of decreasing with time in
a flash. The observed temporal and spectral characteristics give
constraints and valuable hints on the flash production mechanism.
\end{abstract}

\section{Introduction}

   Intense $\gamma$-ray bursts of atmospheric origin were 
detected by the Burst and Transient Source Experiment (BATSE) detectors,
located on the Compton Gamma Ray Observatory (CGRO) 
(\cite{Fishman1994}). All the observed TGFs were of short
duration (a few milliseconds) and had extremely hard energy
spectra consistent with bremsstrahlum radiation from energetic
(MeV) electrons. Since they were spatially correlated with
regions of high thunderstorm activity and in two cases have been
correlated with individual lightning flashes (\cite{inan1996}),
they are thought to be caused by high-altitude discharges
produced by runaway air breakdown. Several theoretical
calculations have been done to explain the origins of TGFs 
under the runaway air breakdown mechanism
(\cite{chan1995}; \cite{tara1996}; \cite{leht1996}; \cite{leht1997}). 
The calculated $\gamma$-ray fluxes and 
duration of the bursts are comparable to those measured by the
BATSE detector on the CGRO satellite.

As the runaway air breakdown is a fundamental new process in plasma,
 it is interesting to understanding more the physics of the process
through studying TGFs, if they were truly caused by such a mechanism.
Besides the total fluxes, durations and
time-averaged spectra, the energy dependence of temporal profiles 
and spectral evolution of TGFs can give more information
about their production process. 
For this purpose we analyze the time-tagged event (TTE) data for TGFs
observed by BATSE. For 4 energy channels with boundaries 25-55
keV, 55-110 keV, 110-320 keV, and $>320$ keV, the TTE data
contains the arrival time ($2\mu s$ resolution) of each photon
from a triggered event up to $\sim 3.3\times 10^4$ photons. Of 47
TGFs observed on CGRO in $\sim 4$ years and published by NASA
in their details, 15 bursts with high signal to noise ratio are used 
in our analysis, which are 
selected by the criterion  of $F_m/T_{90} > 4$ ph/cm$^{-2}$/0.1ms/ms, 
where $F_m$ is the peak rates of the counting series with time 
step 0.1 ms, $T_{90}$ (ms) the burst duration which contains $\% 90$ 
burst photons.

%
%

\section{Energy Dependence of Temporal Profiles}
To study the energy dependence of temporal profiles, which is 
important to diagnose the emission mechanism, we construct 
two counting series $f_1(t)$ and $f_2(t)$  for each burst 
for the energy band 25-110 keV and  $>110$ keV, respectively. 
The average pulse widths in the low and high 
energy bands and the relative time delays between the two bands can be
calculated by a modified correlation analysis technique used in
studying X-ray rapid variability of the black hole binary Cyg X-1
(\cite{Li1999}; \cite{Li2001}).  

The cross-correlation function of  two time
series $f_1$ and $f_2$ at time lag $\tau$ is defined as
\begin{equation}
        \mbox{CCF}(\tau) = \sum_{i}v_1(i\Delta t +\tau)v_2(i\Delta t )/\sigma(v_1)\sigma(v_2)
    \end{equation}
where $v(t) = f(t) - \overline{f}$, $f(t)$ is the number of photons in
 the time interval   $(t, t + \Delta t)$, $\Delta t$ is the time step.
If the function CCF($\tau$)/CCF($0$) has maximum at
$\tau=\Lambda$, the time lag of the energy band 1 relative to the
band 2 at the time scale $Delta t$ is then defined as $\Lambda$.  
Monte Carlo simulations have been done and the results show that with 
this technique we can measure the relative time delay between two bands over 
a wide range of time scale $\Delta t$ with high time resolution.
At large scales one usually can get enough signal photons in a time bin and 
desirable correlation values from finite time bins. 
And at small scales the effect of serious Poisson fluctuation of signal counts 
in a time bin can be compensated by the large amounts of time bin
and accurate correlation values can also be derived with Eq. (1).          
A distribution of time lag vs. time scale can reflect the character of the physical
process to produce the delay better than a single value of lag at only 
one time scale.  
A physical process usually occurs in a range of time scale, the spectral
delay caused by the process should appear at different time scales, smoothly
distributed  in the range. 
On the other hand apparent delays from statistical fluctuation will fluctuate 
between positive and negative values.
  
For each studied burst
and $m=25$ different values of time step $\Delta t$ which are
logarithmically uniformly placed in the region of $10^{-5} -
10^{-3}$ s, we calculate the time lags $\Lambda$ of $f_1$
relative to $f_2$. All obtained time lags $\Lambda_i~ (i=1,...,m)$
of each burst with high signal to noise ratio
are always positive. We average each 5 successive $\Lambda_i$ and show
the distribution of average time lag vs. time scale in Fig. 1.
The global average $\overline{\Lambda}=\sum_{i=1}^{m}\Lambda_i/m$
and the standard deviation
$\sigma(\Lambda)=\sqrt{\sum(\Lambda_i-\overline{\Lambda})^2/(m-1)}$
for each selected TGF is listed in Table 1.  The total counting
profile and profiles in 4 energy bands of a TGF with BATSE
trigger number 2955 are plotted in Figure 2, where 
the delay of lower energy photons relative to higher energy ones
is apparent.

The width $W_l$ of a temporal profile in a band $l$ can be defined as 
the FWHM of the autocorrelation function 
\begin{equation}
       \mbox{ACF}(\tau) = \sum_{i}v_l(i\Delta t +\tau)v_l(i\Delta t )/\sigma^2(v_1)
    \end{equation}
The widths $W_1$ and $W_2$ of studied TGFs in the low and high energy bands
are calculated and their ratios $W_1/W_2$ are presented in Table 1,
from which we can see that
the lower energy pulses are wider than higher energy ones for most studied bursts.

\section{Spectral Evolution}

    For two counts $c_1$ in 25-110 keV band and $c_2$ 
in $>110$ keV band recorded in the same time segment, we 
defined the hardness ratio as
    \begin{equation}
        h = \frac{c_2 - c_1}{c_2 + c_1}
    \end{equation}
To show the spectral evolution,  a TGF's pulse is divided into 5
time segments, each segment contains approximately the same number of 
photons and the hardness ratio is calculated by Eq. (3). Figure 3
shows the spectral hardness variation vs. time (indicating by the
segment number) with four panels grouped by the shape of temporal evolution 
of hardness. Although the shapes are various, a general trend of hardness
that decreases with time appears in Fig. 3, which confirms  
the claim by {\it Nemiroff et al.} [1997] with earlier TGF data.

\section{Discussion}
    For all the bursts with high signal to noise ratio we find that 
in comparison with high
energy band of $>110$ keV, $\gamma$-ray variations in the low
energy band of 25 - 110 keV are always late in the order of 
$\sim 100$ $\mu$s,
pulses are usually wide, and the energy spectra have 
a general trend to
softening with time. The above features of energy dependence of
time profiles and spectral revolution observed in TGFs support
models of runaway breakdown as they are naturally expected for
$\gamma$-rays produced by explosive discharges in plasmas.
Monte Carlo calculations show that large-voltage and high-temperature
pinch plasma columns can generate observed $\gamma$-ray flashes
with energy spectra and spectral evolution characteristics, including
the magnitude of soft lags, consistent with those observed   
in $\gamma$-ray bursts (\cite{Li1997}; \cite{Li1998}). 

Alternative mechanisms have been
proposed to explain the origin of TGFs, e.g. {\it Fargion}
[2001] claimed that TGFs originated from  $10^{15}-10^{17}$ eV
neutrinos of astrophysical nature induced air showers and take
TGFs as first evidences for discovering ultra high energy
(UHE) neutrinos. 
Our Monte-Carlo simulations show that up-going air
showers are hard to produce the observed time lags between hard and
soft energy photons in TGFs. 
In our simulations a geometrical model of the earth, atmosphere and 
BATSE detector is set. High energy $\gamma$-rays initiated by UHE neutrinos 
are produced from the earth's surface, 
following with the up-going air showers. We tracked the secondary 
photons and electrons and record the arriving time and
energy of all $\gamma$-rays reaching the detector, with deposition
energy in the BATSE's region. Simulations were made with
the Monte-Carlo package GEANT3 from CERN. In more than one thousand 
simulated bursts
no time lag between hard and soft photons larger than 10 $\mu$s
has been found.

Quantitative comparison between observed features and expectations 
from models are needed to further confirm the mechanism and to study
physics of the production process.

\begin{acknowledgments}
We thank Prof. Y. Muraki and the reviewers for helpful comments and suggestions.
 This study was supported 
by the Special Funds for Major State Basic Research Projects and 
by the National Natural Science Foundation of China and  made use 
of data obtained through HEASARC Online Service, provided by NASA/GSFC.
\end{acknowledgments}

{}

\hbox{}\vspace{5mm} 
\begin{center}
Table 1 Selected TGFs 
\nopagebreak
\vspace{2mm}
\begin{tabular}{c c c c c c}
\cline{1-6} 
 Trig No. & $T_{90}$ (ms) & $F_m (10^{-3}$ ph/cm$^2$/0.1ms) & 
 $F_m/T_{90}$ & $\overline{\Lambda} \pm \sigma(\Lambda)~(\mu$s)&
 $W_1/W_2$ \\
\cline{1-6}
106&    1.3 &   6.74 &5.19&  102 $\pm$ 28 &	1.19 \\
1433&   1.5 &   7.15 &4.77&   71 $\pm$ 31 &	1.64 \\
2144&   0.7 &   9.62 &13.74& 115 $\pm$ 40 &	1.03 \\
2348&   1.1 &   7.15 &6.50&   88 $\pm$ 29 &	1.04 \\
2370&   0.7 &   8.14&11.63 & 151 $\pm$ 29 &	1.16 \\
2465&   1.0 &   7.89&7.89 & 157 $\pm$ 25 &	1.38 \\
2754&   0.9 &   7.65 &8.50&  120 $\pm$ 29 &	2.10 \\
2808&   1.0 &   5.67 &5.67& 105 $\pm$ 45 &	0.88 \\
2955&   0.7 &   7.89 &11.28&125 $\pm$ 21 &	1.00 \\
3377&   1.0 &   10.11 &10.11& 373 $\pm$ 31 &	1.47 \\
5577&   0.6 &   5.18 &8.63& 81 $\pm$ 33 &	0.91 \\
5587&   1.2 &   6.41 &5.34&   163 $\pm$ 46 &	1.16 \\
5598&   1.1 &   6.17 &5.61&   117 $\pm$ 20 &	0.94 \\
5665&   1.4 &   7.65 &5.46&   265 $\pm$ 43 &	1.54 \\
6773&   0.9 &   3.82&4.25 &   278 $\pm$ 61 &	0.74 \\
\cline{1-6}
\end{tabular}
\end{center}

\begin{figure}
\epsscale{1.5}
\hbox{}\vspace{4cm}
\plotfiddle{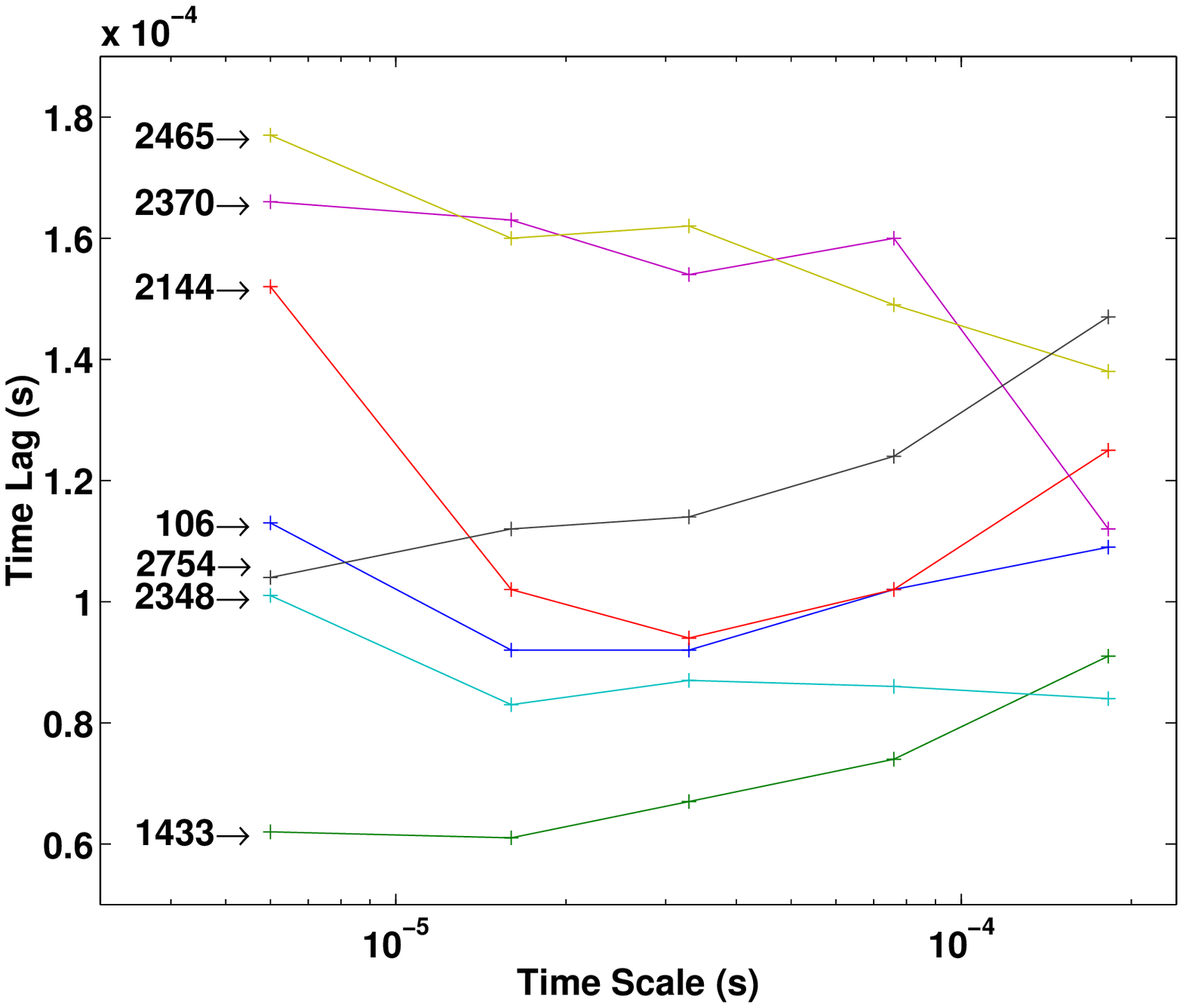}{20pt}{0}{47}{40}{-270}{-100} 
\plotfiddle{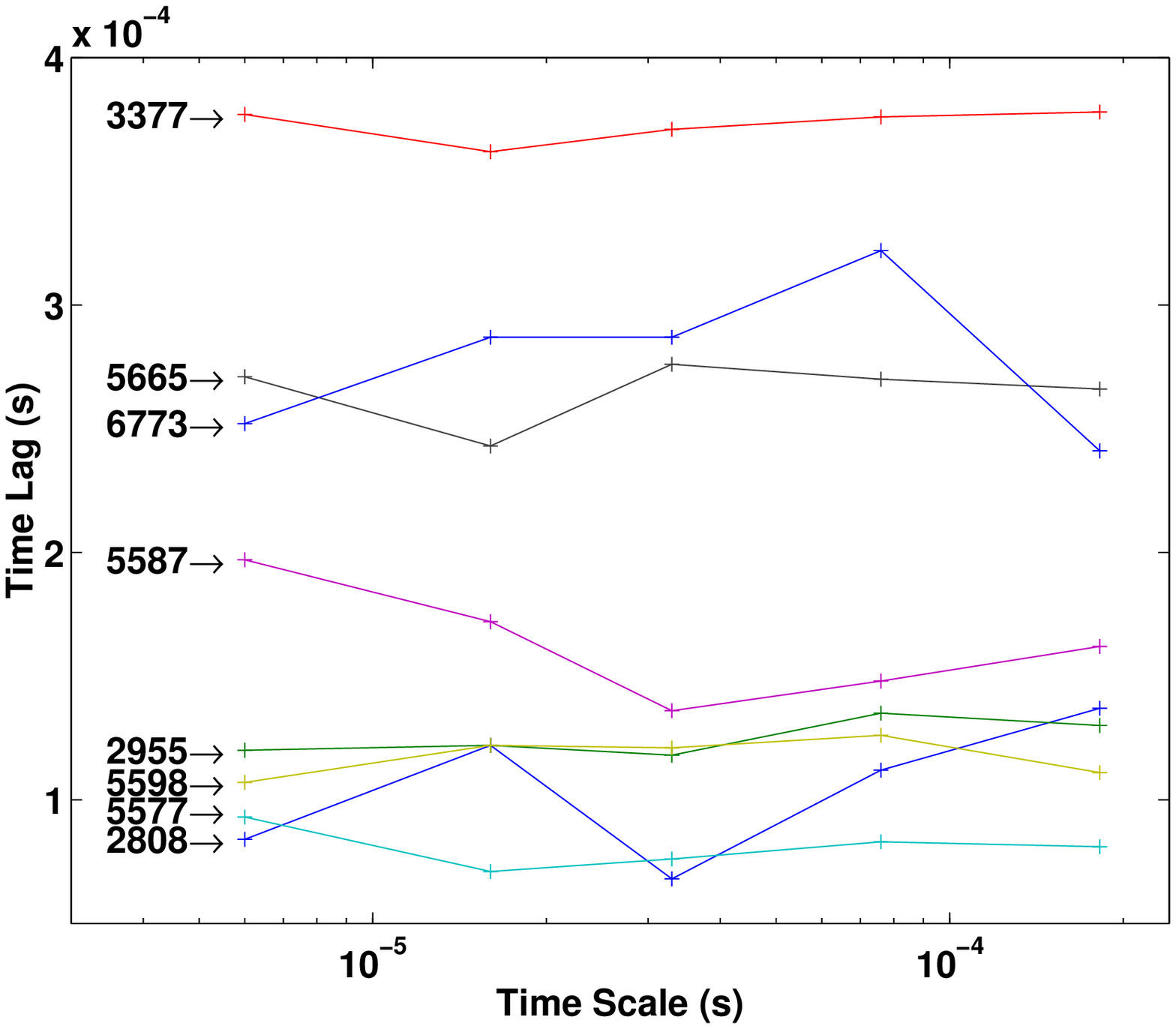}{20pt}{0}{47}{40}{-30}{-65}
\vspace{0.0cm}
\caption{Soft $\gamma$-ray lag vs. time scale of TGFs. The 
quantity beside an arrow is the BATSE trigger 
number of the indicated burst.
\label{fig1}}
\end{figure}

\begin{figure}
\epsscale{1.5}
\hbox{}\vspace{5cm}
\plotfiddle{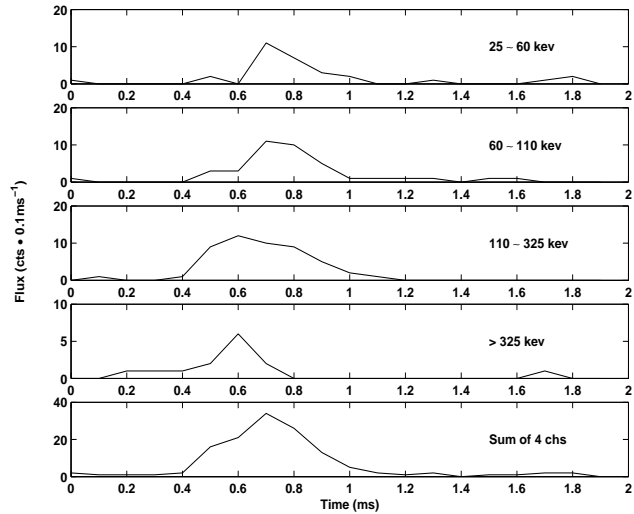}{20pt}{0}{47}{50}{-150}{-100} 
\vspace{0.0cm}
\caption{Counting rate profiles of a TGF with trigger number 2955
in different energy bands. From top to bottom is
the profile in 25-60 keV, 60-110 keV, 110-325 keV, $>325$ keV
and the total BATSE band, respectively.}
\end{figure}

\begin{figure}
\epsscale{1.5}
\hbox{}\vspace{3cm}
\plotfiddle{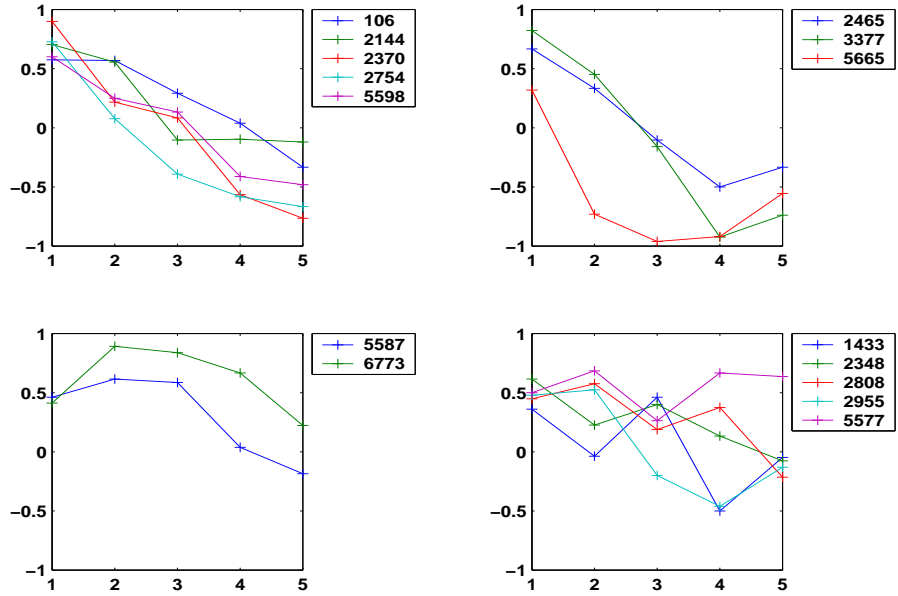}{20pt}{0}{70}{60}{-200}{-100} 
\vspace{0.0cm}
\caption{Spectral evolution of TGFs. The vertical axis is
the hardness ratio defined by Eq. (3). The number marked
along the abscissa is the segment number of a burst profile}
\end{figure}

\end{document}